# Decentralised Service Allocation using Blockchain

Pranshu Chittora

Nikhil Kumar

Ms. G. Malarselvi

# Decentralised Service Allocation using Blockchain

Pranshu Chittora
*Department of Computer Science and Engineering*
*SRM Institute of Science and Technology*
Kattankulathur, India
email: pp5805@srmist.edu.in

Nikhil Kumar
*Department of Computer Science and Engineering*
*SRM Institute of Science and Technology*
Kattankulathur, India
email: nb2556@srmist.edu.in

Ms. G. Malarselvi
*Assistant Professor*
*Department of Computer Science and Engineering*
*SRM Institute of Science and Technology*
Kattankulathur, India
email: malarseg@srmist.edu.in

*Abstract—: "The growing demand for short-term property renting has led to the boom of a new category called PropTech (Property + Technology)". Such a rise in demand attracted many entrepreneurs and investors leading to the inception of powerful and centralised players in the category (like Airbnb).*
*Unfortunately, some big players are controlling the entire industry in a centralised way in turn, performing anti-competitive practices which leave no room for the small players. Such an ecosystem can be made more transparent and decentralised by executing transactions and order-fulfilment on a blockchain.*
*To extend the functionality of the system, we can also perform the renting of small services and appliances via the integration of IoT. Implementing such a system will make the process more transparent, robust and decentralised.*

## I. INTRODUCTION

The growing demand for the rental infrastructure led to the emergence of various centralized players in the ecosystem. But the diverse effect of these major and market-dominating peoples are controlling the majority market in unethical ways.

These centralised entities use the user's data to make sure that the user spends more time by showing him some relevant content. This relevance engine works on a lot of biases. This promotes content moderation and unethical practices. With the proposed system we can replace the centralised control with the decentralised and more trust-based protocol. We can think in terms of migration from Web2 to Web3.

The system will consist of a blockchain-based system with smart contracts for money and asset movements. With the help of running code on the blockchain, the system becomes more transparent and deterministic. Smart contracts can also be used to block money, it's the same as blocking the money on the credit cards. This will allow a security fund kind of system so when something goes wrong or the user of that asset causes some damage then this can be used as a recovery. With asset listing, we can also list services like A/C, washing machine etc as well as some core functionalities like unlocking the door using keys by triggering a function using the owner's key. This can be a game-changer in terms of security and the market potential it comes with.

Transactions on the blockchain are slow and costly so triggering locking and unlocking functionality on the blockchain would be costly and might take a lot of time which is not a good user experience. To mitigate this we can leverage the power of the Ethereum's whisper protocol which is a messaging protocol. This can unlock some huge potential in space.

Blockchain is of two types permissioned and permissionless. The permissionless is more of a public chain where anyone can be the participating node and the trust works on a consensus model. Whereas in a permissioned chain it's more of a private chain and no one can directly join the network with the permission or auth. This is more of a controlled chain and generally used by the government authorities.

## II. PROPOSED WORK

### A. Asset Listing and Allocation

The system will allow users to list their assets (properties) on the main chain so that they can be rented by other user's. A legal owner of the property can list his property by performing a basic KYC and check of the property in order to avoid any listing spam.

Once listed the owner can set the timer period of which he wants to rent the asset. This can be from today to date X or from date X to date Y...

Once committed one can rent the asset from that defined time-bound by executing smart contracts.

### B. Smart Contracts and Messaging Protocol

The system allows temporary change of ownership of the asset on the blockchain this can be achieved by the exchange of assets in this case the currency of the chain with the property. Smart contracts come with deterministic output by executing deterministic functions on the chain. For performing transactions, smart contracts are one of the best ways to do such a thing. But to perform real-time communication we can leverage the messaging protocols that are built right into the core of the system. The whisper protocol is one of the best ways to do real-time P2P messaging.

### C. IMPLEMENTATION

The proposed system is a decentralized web-platform and therefore for the ease of implementation has been divided into separate modules. All of these modules have their own set of individual responsibilities, thus following Separation of Concerns design pattern.

### A. Onboarding

1) User Registration

To enable users smooth and unrestricted access the user must register himself on the public chain. This can be as easy as filling a few details to uploading documents for KYC or compliance reasons.

The user must register and a web server will accept the request and enter the user details in a secure distributed database (like Orbit DB or DGraph). This is great in terms of performance and data redundancy.

2) Generation Public and Private Key

The entire idea of the Blockchain is based on cryptography and it works on asymmetric crypto algorithms. Once the user registers it will get registered on the chain as well which in turn will generate a public and private key.

The public key is the address of the user on the chain which is anonymous whereas the private key is like a password or secret for the data encrypted using the public key.

3) Listing of new Property

A user once complete registration can be able to add assets/properties on the blockchain. This allows users to add/list their assets on the blockchain. This can be done by executing a small smart contract which lists the property and generates a unique address for it.

4) Adding new IoT devices on the network

Another part of the system is connected to IoT devices powered by the blockchain. This is great for renting small appliances or opening door locks automatically.

To connect the device to the blockchain it has to be registered and installed physically. Once installed it has to be provided with a unique id (hash) which makes maps with the user (owner).

B. Transaction

1) Listing of services/assets

As discussed above the system can list properties on the chain but this is not as easy as the conventional systems where this can be achieved by just entering the details in the centralized database. In this case, we are required to make an entry on the blockchain and link that property to the real owner of the property (mapping address to address). This can be done by executing a smart contract which is a small function (code).

2) Allocation and locking resource for a particular user

If a user rents the property for X days the rights will be transferred to the tenant temporarily for X days. Also, an amount needs to be blocked as rent and security. This too can be achieved using the smart contract.

3) Verification and Validation

The listing, delisting, allocation and deallocation must be verified and validated. This is the core of decentralized computing which works on trust and consensus. All transactions will be validated by the nodes in the network and in return a small gas has to be paid as an incentive to be part of the networks.

4) Consistency

The core idea behind the blockchain is consistency which is achieved by immutability of the blocks in the chain. There is no way we can change the data in the blockchain. The reason is behind the infeasibility to break the crypto hash. On the contrary this makes it a bit slow and as of today a public blockchain can handle only thousands of transactions whereas some major payment providers can handle 100 thousands of transactions.

5) Dealing in Cryptocurrency

One of the major functionality of the system is settlements or transfer of funds that has been done from one fiat currency to another fiat currency. This conversion is achieved by converting fiat money to one intermediary form and then to another fiat. This comes with a small cost of payment verification by the consensus and anonymity at its core.

C. IoT

1) Transfer of Ownership

Ownership of a device can be transferred manually or by executing automated smart contracts. Ownership transfer is useful for giving access to an entity to another user.

2) Authentication

Authentication and Authorisation is one of the most important aspects of security. A robust auth is required to safeguard assets from bad actors.

3) Pay for what you use

Pay for what you use is a win-win for all the users as it allows low rental cost allocation of the resources.

4) Automate billing

Automated billing allows hassle free business management. All IoT devices will send the status of usage and aggregation is derived and calculated.

*5) No human intervention required*
All the above can be achieved with no human intervention.

### D. User Management

*1) Manage user data anonymously*
The main idea of the web3 is to maintain privacy at its core. The system allows anonymous storage of user data and minimizing the privacy voiding risk.

*2) Handle asset listing and delisting*
Handling assets carefully is critical for the entire system in terms of technical as well as legal aspects. A user can change the ownership of the property in the real world which must reflect in the blockchain. This is crucial so that the rent must be paid to the real owner not the previous owner of the property. This can be resolved by performing regular check or connecting to the land record department directly in the future.

*3) User engagements/activities*
A user can engage with the system in various manners. A user doesn't want to see a property which he is not interested in due to geographical reasons. Therefore a user can apply various filters and constraints to reduce the search space. This can be a geographical filter or a user preference filter.

### E. Legal Documents

*1) Legal Compliances*
The system revolves around allocating physical/real assets in the virtual world therefore it is quite necessary to verify the actual relationship of the user with that asset. Therefore localized compliance policies must imply.

*2) User KYC for sensitive assets*
The above module can be extended to KYCs and collection of docs.

*3) Privacy Policies*
The system focuses on privacy at its core therefore strict privacy policies must be implied, in order to protect the user data and maintain anonymity .